\documentclass{JHEP3} 

\pdfoutput=1

\usepackage{amsmath}

\usepackage{amssymb}

\usepackage{epsf}

\usepackage{graphicx}

\usepackage{enumerate}

\newcommand{\be}{\begin{equation}}
\newcommand{\ee}{\end{equation}}
\newcommand{\beq}{\begin{equation}}
\newcommand{\eeq}{\end{equation}}
\newcommand{\bea}{\begin{eqnarray}}
\newcommand{\eea}{\end{eqnarray}}

\newcommand{\ms}{\overline{\mathrm{MS}}}

\newcommand{\sumint}[1]{\hbox{$\sum$}\!\!\!\!\!\!\!\int_{#1}}

\title{Thermodynamics and phase diagram of anisotropic Chern-Simons deformed gauge theories}

\author{Antti Gynther, Anton Rebhan and Dominik Steineder\\
  Institut f\"{u}r Theoretische Physik, 
  Technische Universit\"{a}t Wien,\\
  Wiedner Hauptstr. 8--10,
  1040 Vienna, Austria\\ 

  \\

  \email{gynthera, rebhana, steineder @hep.itp.tuwien.ac.at}
}

\abstract{We consider 3+1-dimensional gauge theories at finite temperature and a finite density of charges which couple to a 2+1-dimensional Chern-Simons operator, giving rise to a $\theta$-term with constant spatial gradient of $\theta$. The strong-coupling limit of thermal $\mathcal N=4$ super-Yang-Mills theory with this kind of anisotropic deformation has been used in the context of the AdS/CFT correspondence as a model for strongly coupled anisotropic quark-gluon plasma. In this paper we work out the thermodynamics and the (nontrivial) phase diagram in the limit of vanishing gauge coupling and compare with the corresponding strong-coupling results.}
\begin{document}

\maketitle

\section{Introduction}

There is by now little doubt that quark-gluon plasma can be produced and studied in current ultrarelativistic heavy-ion collider experiments (RHIC at BNL, LHC at CERN) \cite{Muller:2006ee}. The quark-gluon plasma created in this way is however initially very far from equilibrium, and it may have substantial pressure anisotropies over its entire lifetime until freeze-out \cite{Florkowski:2010cf,Martinez:2009ry,Martinez:2010sd,Martinez:2010sc}.

At weak coupling, an anisotropic plasma exhibits plasma instabilities \cite{Weibel:1959,Mrowczynski:1988dz,Romatschke:2003ms,Arnold:2003rq,Rebhan:2009ku} which in the non-Abelian case have complicated dynamics that has been studied extensively by numerical approaches using the approximation of stationary anisotropy \cite{Rebhan:2004ur,Arnold:2005vb,Rebhan:2005re,Arnold:2005ef,Arnold:2007cg,Bodeker:2007fw,Ipp:2010uy}, and more recently also for anisotropic expansion \cite{Romatschke:2006wg,Rebhan:2008uj,Attems:2012}.\footnote{Similar instabilities have been identified in the so-called glasma phase, where the dynamics of non-Abelian gauge fields after the collision of ultrarelativistic color sources is governed in leading order of perturbative QCD by classical Yang-Mills field equations  \cite{Romatschke:2006nk,Fukushima:2011nq} as well as in classical-statistical simulations of Yang-Mills field dynamics \cite{Berges:2008mr,Berges:2008zt}.} These plasma instabilities are crucial for understanding thermalization and isotropization of a weakly coupled plasma \cite{Arnold:2003rq,Kurkela:2011ti,Kurkela:2011ub}. Moreover, they lead to ''anomalously'' low effective viscosity and could thus mimic an inherently strong-coupling situation \cite{Asakawa:2006tc}. It is therefore of considerable interest to develop a more complete understanding of the effects of anisotropies in both weakly and strongly coupled plasmas.

At strong coupling, where the AdS/CFT correspondence \cite{Aharony:1999ti} has provided new tools \cite{CasalderreySolana:2011us}, the effects of (temporarily fixed) anisotropy in a supersymmetric Yang-Mills plasma have been modeled by singular geometries involving comparatively benign naked singularities \cite{Janik:2008tc,Rebhan:2011ke} and more recently by a completely regular construction involving axion-dilaton gravity \cite{Mateos:2011ix,Mateos:2011tv}. 
Interestingly enough, this latter model has been found to contain phases with instabilities reminiscent of the filamentation instabilities at weak coupling. In follow-up works, several observables of interest to heavy-ion physics have been studied in this model \cite{Chernicoff:2012iq,Giataganas:2012zy,Chernicoff:2012gu,Rebhan:2012bw,Fadafan:2012qu} and compared with calculations for a weakly coupled anisotropic plasma \cite{Romatschke:2006bb,Schenke:2006yp,Dumitru:2007hy,Baier:2008js,Dumitru:2009ni,Burnier:2009yu,Philipsen:2009wg}, a perhaps particularly remarkable finding being that of a shear viscosity coefficient breaking the Kovtun-Son-Starinets bound \cite{Rebhan:2011vd,Mamo}.\footnote{See \cite{Polchinski:2012nh} for a related finding in a different gauge-string dual.}

While other holographic models for anisotropic fluids have been constructed \cite{Erdmenger:2011tj,Oh:2012zu,Gahramanov:2012wz}, the model of Ref.~\cite{Mateos:2011ix,Mateos:2011tv}, which builds upon the string theory dual constructed in Ref.~\cite{Azeyanagi:2009pr}, is especially attractive because it is a string-theoretic top-down construction. The stationary anisotropy is brought about by an anisotropic distribution of D7 branes which apart from wrapping the $S^5$ of the bulk geometry fill only 2 out of the 3 spatial dimensions. Like the color D3 branes they do not extend along the holographic direction, but are dissolved in the geometry. The corresponding gravitational background involves an axion with constant spacelike gradient. 

On the gauge theory side, this corresponds to a deformation of $\mathcal N=4$ super-Yang-Mills theory by a position-dependent $\theta$-term
\be
\delta S=\int \frac{n_{D7}z}{4\pi} \,{\rm Tr}\, F \wedge F
=\int \frac{az}{g_{\rm YM}^2}  \,{\rm Tr}\, F \wedge F
\ee
where $z$ is the spatial coordinate along which a constant density $n_{D7}$ of D7 branes is set up.
Absorbing $g_{\rm YM}$ in $F$ by a rescaling this leads to the modified Yang-Mills Lagrangian
\be\label{Lanisoqcd}
\mathcal L=-\frac14 F_{\mu\nu}^a F^{a\mu\nu}-\frac{1}4 
\tilde\theta(x) \epsilon^{\mu\nu\rho\sigma}F_{\mu\nu}^a F^a_{\rho\sigma}
\ee
with $\tilde\theta(x)=az$. By partial integration one finds that the charge density $a$ is coupled to a homogeneous but anisotropic operator given by a 2+1-dimensional Chern-Simons term.

The aim of the present paper is to study the thermodynamics and phase diagram of this anisotropic Chern-Simons deformed gauge theory at vanishing coupling and to compare with the corresponding results obtained by means of gauge-gravity duality in Ref.~\cite{Mateos:2011tv}. As we shall see, the phase diagram at zero coupling is even richer than the one at strong coupling, with significant differences in particular at high temperatures, $T\gg a$.

A similar Chern-Simons deformation of electrodynamics has originally been studied by Carroll, Field, and Jackiw in \cite{Carroll:1989vb} as a model for a Lorentz and CPT violating electrodynamics which preserves rotational invariance by a timelike gradient of the $\theta$ parameter. In this version of the theory there are tachyonic modes, which are however absent in our case of interest, the case of a spacelike $\theta$ gradient \cite{Ralston:1995rt,Adam:2001ma} (which for a while has attracted attention as a possible explanation for cosmic anisotropy in the polarization of distant radio sources \cite{Nodland:1997cc} but evidence for the latter was refuted by Carroll and Field \cite{Carroll:1997tc,Jackiw:1998js}).

As we shall discuss further below, the photons in this modified anisotropic electrodynamics have dispersion laws of the form
\begin{equation}
  \omega_\pm^2=\mathbf k^2+\frac{a^2}2\left(
  1\pm \sqrt{1+\frac{4 k_\parallel^2}{a^2}} \right).
\end{equation}
There are gauge boson modes with a mass gap $a$ as well as ungapped ones,
but whenever there is a wave vector component parallel to the direction of anisotropy, there is a deviation from an ordinary mass shell or the light-cone. While ungapped modes have $\omega_- \le |\mathbf k|$, all modes have $\omega_\pm^2\ge0$ and are therefore stable, which is in fact markedly different from the gauge boson propagator in a weakly coupled anisotropic plasma, where a rich spectrum of instabilities and unstable modes arises \cite{Romatschke:2003ms,Arnold:2003rq}. Nevertheless, like in the strong coupling case we shall identify phases with thermodynamic instabilities against inhomogeneous redistribution of the ``Chern-Simons charge'' density $a$.

\section{Setup and notation}
\label{sec-1}

We consider a theory of free photons in a system containing a source $j(x)$ for 2+1 dimensional Chern-Simons operator. The partition function is given by the Euclidean path integral (in Euclidean spacetime, we write all the Lorentz indices as subscripts to distinguish it from Minkowski spacetime)
\begin{eqnarray}\label{pathint}
  \mathcal{Z}(T,j) & = & \int \mathcal{D}A_\mu \mathrm{exp}\left[-\int_0^\beta d\tau\int d^3x \left(\mathcal{L} + \frac{i}{4} j(x)\epsilon_{\mu\nu\rho\sigma}A_\mu F_{\nu\rho}\zeta_\sigma + \Omega\right) \right], \\
  \mathcal{L} & = & \frac{1}{4}F_{\mu\nu}F_{\mu\nu} + \frac{1}{2}\left(\partial_\mu A_\mu \right) + \mathrm{ghosts},
\end{eqnarray}
where the constant spacelike unit vector $\zeta_\mu$ specifies the 2+1 dimensional subspace, the factor $1/4$ is a convention, $\Omega$ is the cosmological constant (needed for renormalization) and we have adopted the Feynman gauge. Note that the charges that couple to $\epsilon_{\mu\nu\rho\sigma}A_\mu F_{\nu\rho}\zeta_\sigma$ 
can be localized only in the $\zeta$-direction, $j(x) = h(x_\mu \zeta_\mu)$, otherwise gauge invariance is broken. We consider a system in which we have a constant charge density along the $\zeta$-direction (which without loss of generality can be taken as the $z$-direction),
\begin{eqnarray}
  j(z) & = & a \equiv \frac{N}{L_\parallel},
\end{eqnarray}
so that (\ref{pathint}) is anisotropic but translationally invariant.
Here, $N$ is the number of charges coupling to the Chern-Simons operator and $L_\parallel$ the extent of the system in $\zeta$-direction. In order to make contact with the notation of Ref.~\cite{Mateos:2011ix,Mateos:2011tv}, we write for the constant density $j(z)=a$. Denoting the linear extent of the system perpendicular to $\zeta$-direction as $L_\perp$, the volume is then $V=L_\perp^2L_\parallel$.

We use the notation $K^2 = k_\mu k_\mu = k_0^2 + \mathbf k^2 = k_0^2 + k_\perp^2 + k_\parallel^2$ for momenta. We renormalize the theory in dimensional regularization by splitting the spacetime into $\mathbf{R}^3\times\mathbf{R}^{1-2\epsilon}$ (``transverse'' $\times$ ``longitudinal''), with the
Chern-Simons term taking the form $\epsilon_{ijk}A_iF_{jk}$ with $\{i,j,k\}$ labeling the transverse directions (including Euclidean time). Additionally, we label the longitudinal directions with letters from the beginning of the Greek alphabet, $\alpha,\beta,\dots$, leaving
$\mu,\nu,\dots$ to label directions in the entire spacetime.

Our notation for momentum integrations is such that
\begin{eqnarray}
  \int_k & = & \int_{k_\parallel} \int \frac{d^2k_\perp}{(2\pi)^2} \, = \, \left(\frac{e^\gamma\bar{\Lambda}^2}{4\pi}\right)^\epsilon\int\frac{d^{1-2\epsilon}k_\parallel}{(2\pi)^{1-2\epsilon}}\frac{d^2k_\perp}{(2\pi)^2}, \\
  \sumint{K} & = & T\sum_{k_0}\int_k 
\end{eqnarray}

\section{Thermodynamic functions}
\label{sec-2}

Before carrying out explicit calculations, we discuss the various thermodynamic functions of interest to us (see also Appendix C of Ref.~\cite{Mateos:2011tv}).

Energy is a function of extensive variables, $E=E(S,N,L_\perp,L_\parallel)$. Pressures in various directions are most naturally expressed in terms of it, and are given by
\begin{eqnarray}\label{eq:pres.defs}
  p_\perp & = & -\frac{1}{L_\parallel}\left(\frac{\partial E}{\partial L_\perp^2}\right)_{S,N,L_\parallel},\quad p_\parallel \, = \, -\frac{1}{L_\perp^2}\left(\frac{\partial E}{\partial L_\parallel}\right)_{S,N,L_\perp}.
\end{eqnarray}
A more convenient quantity to compute, however, is the Helmholtz free energy, $F = F(T,N,L_\perp,L_\parallel) \equiv E - TS$, given by
\begin{eqnarray}
  F & = & -T \ln \mathcal{Z}\, = \, V f(T,N/L_\parallel) \,=\, V f(T,a).
\end{eqnarray}
(The functional form $F(T,N,L_\perp,L_\parallel)=Vf(T,N/L_\parallel)$ is clear from the form of the partition function.)
From the definitions of pressures in Eq. (\ref{eq:pres.defs}) and the relation between $F$ and $E$, we immediately find that the pressures are given by
\begin{eqnarray}
  p_\perp & = & -\frac{1}{L_\parallel}\left(\frac{\partial F}{\partial L_\perp^2}\right)_{T,N,L_\parallel} \, = \, -\frac{1}{L_\parallel}\left(\frac{\partial F}{\partial L_\perp^2}\right)_{T,a} \, = \, -f \\
  p_\parallel & = & -\frac{1}{L_\perp^2}\left(\frac{\partial F}{\partial L_\parallel}\right)_{T,N,L_\perp} \, = \, -f - L_\parallel\left(\frac{\partial f}{\partial L_\parallel}\right)_{T,N} = -f + a\left(\frac{\partial f}{\partial a}\right)_T.
\label{pparallel}
\end{eqnarray}

Note that even though our physical system is inherently anisotropic as soon as $a\not=0$ (with, e.g., anisotropic relations between energies and momenta of its particles, see below), the pressure would be necessarily isotropic in thermal equilibrium if $a$ did not react differently to changes in the system size along different directions.

We can also study the system in the grand canonical ensemble, as a function of the chemical potential $\mu \equiv \partial F/\partial N$ conjugate to the number $N$ of the Chern-Simons charges. The associated free energy,
the grand potential, is given by
\begin{eqnarray}
  G & \equiv & F - N\left(\frac{\partial F}{\partial N}\right)_{T,L_\perp,L_\parallel} \, = \, V\left[f - a\left(\frac{\partial f}{\partial a}\right)_T\right].
\end{eqnarray}
The grand potential has the general form $G(T,\mu,L_\perp,L_\parallel)=Vg(T,\mu/L_\perp^2)$, which is easy to see from above as follows. Defining 
\be
\Phi \equiv \frac{\partial f}{\partial a},
\ee
 it is clear by construction that the grand potential
can be written as 
\be
G = V g(T,\Phi)
\quad \text{with} \quad
g(T,\Phi) = f - a\frac{\partial f}{\partial a}. 
\ee
On the other hand, we have
\begin{eqnarray}
  \mu & \equiv & V\left(\frac{\partial f}{\partial N}\right)_{T,L} = L_\perp^2\left(\frac{\partial f}{\partial a}\right)_{T} = L_\perp^2\Phi,
\end{eqnarray} 
and thus $\Phi = \mu/L_\perp^2$ and $G=Vg(T,\mu/L_\perp^2)$. In terms of the grand potential, the pressures can then be written as
\begin{eqnarray}
  p_\perp & = & -\frac{1}{L_\parallel}\left(\frac{\partial G}{\partial L_\perp^2}\right)_{T,\mu,L_\parallel} \, = \, -g - L_\perp^2\left(\frac{\partial g}{\partial L_\perp^2}\right)_{T,\mu} = -g + \Phi\left(\frac{\partial g}{\partial \Phi}\right)_T, \\
  p_\parallel & = & -\frac{1}{L_\perp^2}\left(\frac{\partial G}{\partial L_\parallel}\right)_{T,\mu,L_\perp} \, = \, -\frac{1}{L_\perp^2}\left(\frac{\partial G}{\partial L_\parallel}\right)_{T,\Phi} \, = \, -g\,.
\end{eqnarray}

Instead of working with $F$ and $G$, which are functions of conjugate variables $N$ and $\mu$, respectively,
it is more convenient to consider the free energy densities $f$ and $g$, which are functions of the conjugate variables $a$ (number of charges per unit length) and $\Phi$ (chemical potential per unit transverse area), respectively.

\section{Results at vanishing gauge coupling}
\label{sec-3}

\subsection{Free energy}
\label{sec-3.1}

Since we are considering a free theory, the path integral in Eq.~(\ref{pathint}) for a homogeneous source $j(z) = a$ can be carried out using standard methods. Details of all the computations are given in Appendix \ref{app:sec-1}; we will quote only the essential results here. The resulting Helmholtz free energy density $f(T,a) = -T/V \ln\mathcal{Z}(T,a)$ is given by
\begin{eqnarray}\label{free-en}
  f(T,a) & = & \Omega + T\sum_{\pm}\int_k\left[\frac{1}{2}\beta\omega_\pm + \ln\left(1-e^{-\beta\omega_\pm}\right)\right],
\end{eqnarray}
where
\begin{eqnarray}
  \omega_\pm^2 & = & k_\perp^2 + M_\pm^2(k_\parallel), \quad M_\pm^2(k_\parallel) \, = \, k_\parallel^2 + \frac{a^2 \pm \sqrt{a^4+4a^2 k_\parallel^2}}{2},
\end{eqnarray}
or, equivalently,
\be
M_\pm(k_\parallel) 
=  \sqrt{k_\parallel^2 + a^2/4} \pm \frac{a}{2}. 
\ee
Note that (\ref{free-en}) is an even function of $a$ so that without loss of generality we can assume $a\ge0$.

The $T=0$ limit of Eq.~(\ref{free-en}) contains a UV-divergence that needs to be renormalized. In the $\ms$-scheme, we obtain
\begin{eqnarray}\label{eq:f0}
  f(0,a) & = & -\frac{c\,a^4}{12\pi^2} + \frac{5 a^4}{256 \pi^2}\ln\frac{a}{\bar{\Lambda}} + \Omega(\bar{\Lambda}),
\end{eqnarray}
where
\begin{eqnarray}
  c & = & \int_0^\infty dx \left[\left(x^2+\frac{1}{2}(1+\sqrt{1+4x^2})\right)^{3/2}+\left(x^2+\frac{1}{2}(1-\sqrt{1+4x^2})\right)^{3/2}\right. \nonumber \\
    & & \quad \quad \left.-2 x^3 - \frac{9}{4}x - \frac{15}{64\sqrt{1+x^2}}\right] = 0.29136\ldots  
\end{eqnarray}
and $\Omega(\bar{\Lambda})$ is the renormalized cosmological constant running with the scale $\bar{\Lambda}$. Note that Eq.~(\ref{eq:f0}) yields the exact renormalization group equation governing the running of the cosmological constant in this model. However, since we are primarily interested in the thermodynamics as a function of $a$ at some fixed (but arbitrary) scale $\bar{\Lambda}$, we are free to choose the value of the cosmological constant (which by definition is independent of $a$) at that scale. With this in mind, fixing $\bar{\Lambda}$ to give units $a$ is measured in, we set $\Omega(\bar{\Lambda})=0$, yielding for the $T=0$ free energy
\begin{eqnarray}
  f(0,a) & = & -\frac{c\,a^4}{12\pi^2} + \frac{5 a^4}{256 \pi^2}\ln\frac{a}{\bar{\Lambda}}.
\end{eqnarray}
The coefficient in front of the logarithm gives the trace anomaly of our system,
\be\label{tranom0}
\epsilon-2 p_\perp-p_\parallel = -\frac{5 a^4}{256 \pi^2}\,.
\ee

Note that the free energy vanishes at a finite value of the density,\footnote{In case we had not set the cosmological constant to zero, the equivalent statement is that the free energy has the same value at some finite density $a=a_0$ as at $a=0$.} $a=a_0$, and we can thus express the scale $\bar{\Lambda}$ in terms of $a_0$,
\begin{eqnarray}
  \ln\frac{a_0}{\bar{\Lambda}} & = & \frac{64c}{15}
\end{eqnarray}
such that
\begin{equation}\label{free-en0}
f(0,a) \, = \, \frac{5 a^4}{256 \pi^2}\ln\frac{a}{a_0}.
\end{equation} 
From now on, we express everything in units of $a_0$.

Other thermodynamic variables of interest at $T=0$ can now be computed,
\begin{eqnarray}
  \Phi \, = \, \frac{\partial f}{\partial a} & = & \frac{5 a^3}{256 \pi^2}\left(1+4\ln\frac{a}{a_0}\right), \\
  p_\parallel \, = \, -f + a\Phi & = & \frac{5 a^4}{256 \pi^2}\left(1+3\ln\frac{a}{a_0}\right), \\
  \Phi' \, = \, \frac{\partial \Phi}{\partial a} & = &  \frac{5 a^2}{256 \pi^2}\left(7+12\ln\frac{a}{a_0}\right).
\end{eqnarray} 
Respectively, they vanish at different densities $a$ given by
\begin{eqnarray}
\label{aPhi0}
  a_\Phi & = & a_0\, e^{-1/4} \, \approx \, 0.7788\, a_0, \\
\label{app0}
  a_{p_\parallel} & = & a_0\, e^{-1/3} \, \approx \, 0.7165\, a_0, \\
\label{aPhip0}
  a_{\Phi'} & = & a_0\, e^{-7/12} \, \approx \, 0.5580\, a_0.
\end{eqnarray}

The finite-$T$ contribution is given by
\begin{eqnarray}
  f(T,a)-f(0,a) & = & T\sum_{\pm}\int_k\ln\left(1-e^{-\beta \omega_\pm}\right).
\end{eqnarray}
Unfortunately, it is not possible to express this sum-integral in a closed form. We can, however, derive a simpler integral representation for the finite-$T$ contribution that is straightforward to evaluate numerically (see Appendix \ref{app:sec-1.2} for details),
\begin{equation}\label{free-enT}
    f(T,a)-f(0,a) \, = \, \frac{T^4}{2\pi^2}\sum_{\pm}\int_0^\infty dx\, x^2\left(1 \pm \frac{y}{\sqrt{x^2+y^2}}\right)\ln\left(1-e^{-\sqrt{x^2+y^2}\, \mp y}\right)\,,
\end{equation}
where $y=a/(2T)$. 
The integral in (\ref{free-enT}) is of the same form as integrals encountered in the thermodynamics of bosons of mass $a/2$ and chemical potential $-a/2$ (specifically, the first term gives exactly the free energy of such bosons), and we can use standard methods to retrieve high- and low temperature expansions of it. At high temperatures ($T\gg a$) we obtain
\begin{equation}\label{fhiT0}
    f(T,a)-f(0,a) \, = \, -\frac{\pi^2 T^4}{45} + \frac{a^2 T^2}{48} - \frac{a^3 T}{64} - \frac{5 a^4}{256\pi^2}\left(\ln\frac{a}{8\pi T} + \gamma_E - \frac{1}{60} \right) + \mathcal{O}(a^6),
\end{equation}
while at low temperatures ($T\ll a$) we get
\begin{equation}\label{flowT0}
    f(T,a)-f(0,a) \, = \, - \frac{3\zeta(7/2)}{8\pi^{3/2}}\,a^{1/2}T^{7/2} + \mathcal{O}(a^{-1/2}).
\end{equation}
These expansions are useful in studying the asymptotic behaviour of the system, but in the following analysis we solve the system numerically using the exact results in Eqs.~(\ref{free-enT}) and (\ref{free-en0}).

\subsection{Phase diagram and instabilities}
\label{sec-3.3}

The system we consider contains a fixed  number $N$ of charges that couple to the Chern-Simons term. However, how these charges are distributed along the $z$-axis is a free parameter determined by minimizing the free energy $F$. 
We refer to these various distributions as various phases of the system. For simplicity, we only consider distributions which consist of homogeneous regions of finite extent along the direction of anisotropy with various densities, since we know how to compute the free energy density of such homogeneous regions. We refer to the phase with just one region as homogeneous phase, and the phase with two or more homogeneous regions as inhomogeneous phase. Note that the boundaries between regions in the inhomogeneous phase must be perpendicular to the $z$-axis, since gauge invariance requires $a$ to be just a function of $z$.
Hence, any resulting inhomogeneous structure is ``lasagne-like''.

In cases where the homogeneous phase is energetically disfavored, it can still be metastable, which means that finite charge fluctuations are required for the transition to the inhomogeneous phase. When arbitrarily small charge fluctuations are sufficient for lowering the free energy, the system is called thermodynamically unstable.

\subsubsection{Coexistence of homogeneous regions}
\label{sec-3.3.1}

In the inhomogeneous phase, the free energy can be minimized by
a redistribution of the given number of charges into an arbitrary
number of homogeneous regions. We can analyse this situation by
considering the case of two neighboring homogeneous regions.

Two separately homogeneous regions can coexist if the intensive thermodynamic variables have the same values in each region. In the present case this means regions with different values of $a$ can coexist if the longitudinal pressure $p_\parallel$ has the same value in each region (``mechanical equilibrium''). Since the boundary between regions must be perpendicular to the $z$-axis, we do not need to require equality of the transverse pressures. In addition, also the chemical potential $\Phi$ must be the same in each region (``chemical equilibrium''), as long as each region contains a non-zero number of charges so that processes taking charges from one region to another can remain in equilibrium. If one region is empty of charges (and hence isotropic), chemical equilibrium cannot be reached.\footnote{Charges that couple to the Chern-Simons term are not dynamical in the model, which means that formation of charge-anticharge pairs is excluded. Hence, regions with negative charge density cannot emerge.}

\begin{figure}[htb]
  \centering
  \includegraphics[width=10cm]{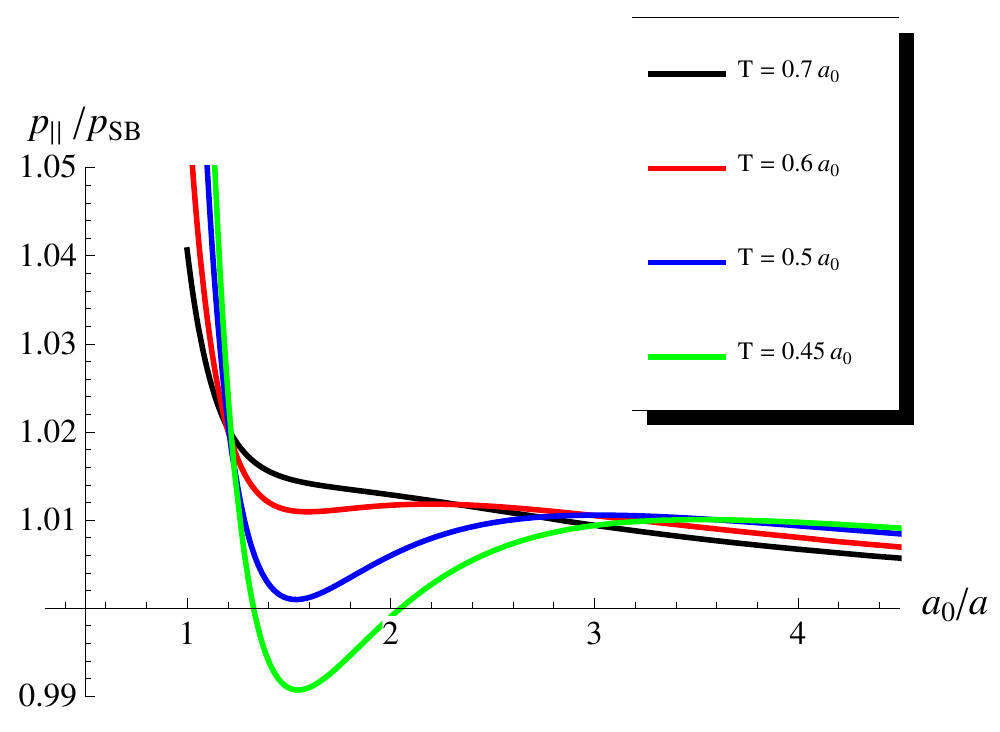}
  \caption{\label{pLplot}Longitudinal pressure as a function of $1/a$ for various temperatures, normalized to Stefan-Boltzmann pressure.}
\end{figure}

In Fig.~\ref{pLplot} we have plotted the longitudinal pressure as a function of $1/a$ ($=L_\parallel/N$, so this corresponds to the traditional $p$-$V$ diagram) for various temperatures. As can be seen, for sufficiently low temperatures, there are various values of $a$ that yield the same longitudinal pressure. Moreover, it can immediately be seen that the homogeneous phase is thermodynamically unstable for certain values of $a$, since the condition
\begin{eqnarray}
  \left(\frac{\partial p_\parallel}{\partial L_\parallel}\right)_{T,N} & < & 0\quad \Rightarrow \left(\frac{\partial p_\parallel}{\partial (1/a)}\right)_{T} \, < \, 0,
\end{eqnarray}
would be violated. Equivalently, we can express this condition as
\begin{eqnarray}
  \left(\frac{\partial^2 f}{\partial a^2}\right)_T = \left(\frac{\partial \Phi}{\partial a}\right)_T & > & 0.
\end{eqnarray}
A homogeneous phase violating this condition cannot exist in equilibrium. We can thus conclude that, for sufficiently low temperatures and certain values of the overall density $a$, the system must reside in the inhomogeneous phase. On the other hand, for sufficiently large temperatures, the system must reside in the homogeneous phase since the longitudinal pressure is then a monotonically increasing function of $a$.

More precisely, the conditions for coexistence of two homogeneous regions with densities $a_1$ and $a_2$, $0 \leq a_1 < a_2$, are given by:
\begin{enumerate}[(a)]
\item coexistence of two anisotropic regions $(a_1 > 0)$:
\end{enumerate}
\begin{eqnarray}
  \label{conds1}
  \Phi(T,a_1) \,\, = \,\, \Phi(T,a_2) & \equiv & \Phi \\
  \label{conds2}
  \frac{f(T,a_2)-f(T,a_1)}{a_2-a_1} & = & \Phi
\end{eqnarray}
\begin{enumerate}[(b)]
\item coexistence of an isotropic and an anisotropic region $(a_1 = 0)$:
\end{enumerate}
\begin{eqnarray}
  \label{conds3}
  f(T,a_2) - a_2\Phi(T,a_2) & = & f(T,0) \, = \, -\frac{\pi^2T^4}{45}.
\end{eqnarray}

\begin{figure}[htb]
  \centering
  \includegraphics[width=7.5cm]{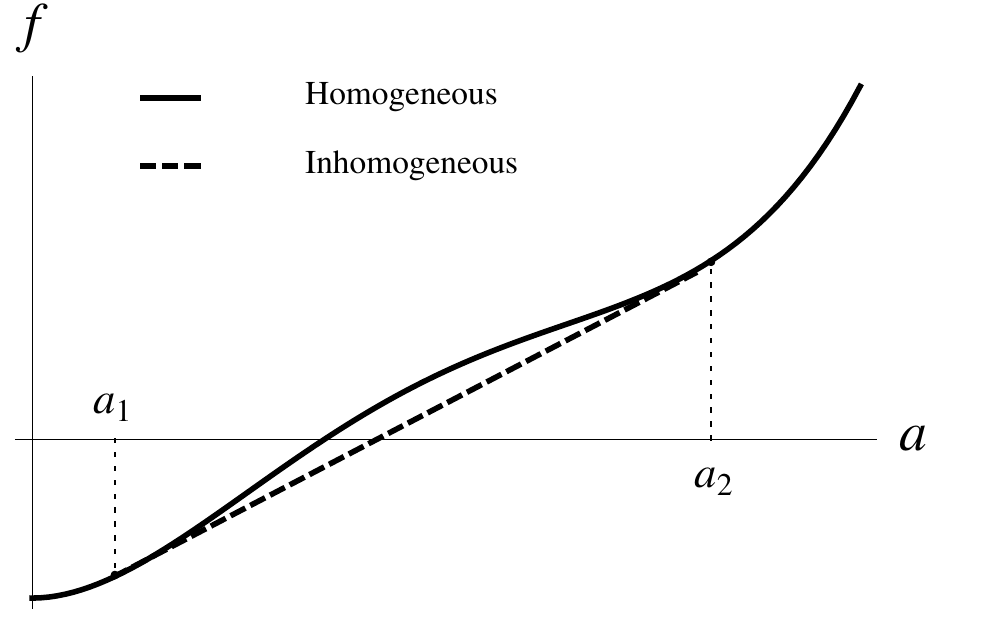}
  \includegraphics[width=7.5cm]{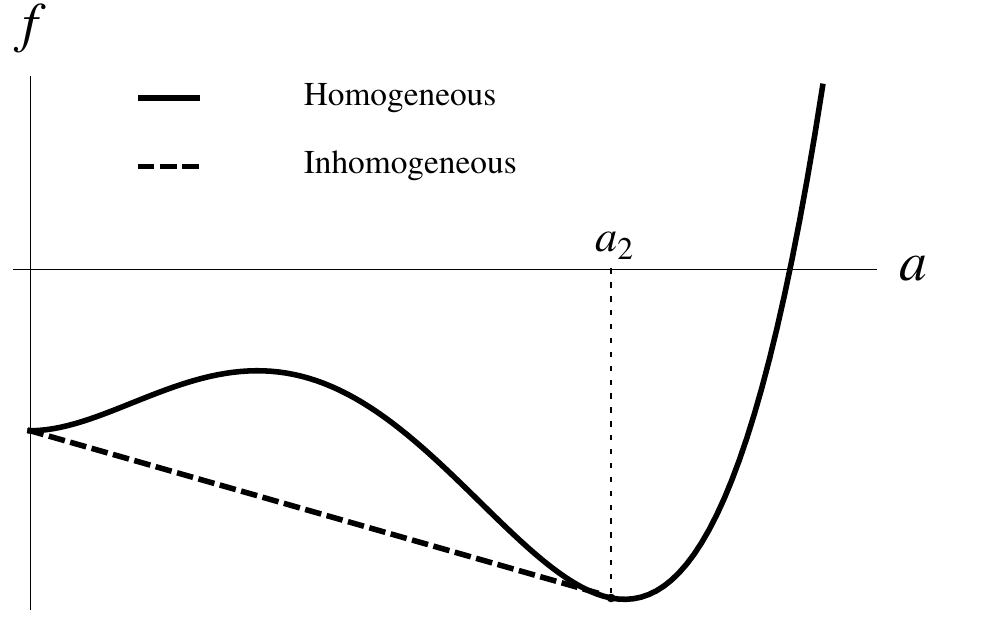}
\centerline{\hfil (a) \hfil\hfil\hfil (b) \hfil}
  \caption{\label{coexist}Two cases where the conditions for coexistence of homogeneous regions are realized. (a)~The dashed line, giving the free energy of the inhomogeneous phase, is the tangent to $f(T,a)$ at points $a=a_1$ and $a=a_2$. (b) The dashed line is the tangent to $f(T,a)$ at $a=a_2$ and coincides with $f(T,a)$ at $a=0$.}
\end{figure}

These two cases are graphically represented in Fig.~\ref{coexist}. Two anisotropic regions with densities $a_1$ and $a_2$ can coexist if the straight line drawn through points $(a_1,f(T,a_1))$ and $(a_2,f(T,a_2))$ is the tangent to the $f(T,a)$ curve at the points $a_1$ and $a_2$ (Fig.~\ref{coexist}a). That straight line corresponds to the free energy of a system containing a mixture of homogeneous phases with densities $a_1$ and $a_2$, that is, the free energy of the inhomogeneous phase. If that line lies below the $f(T,a)$ curve, then the free energy of the inhomogeneous phase is lower than that of homogeneous phase with corresponding overall density, and thus the equilibrium state is given by the inhomogeneous phase. Since stability of the homogeneous regions requires that $\partial^2 f/\partial a^2|_{a_1,a_2} > 0$, at least part of the $f(T,a)$ curve must lie above the line. On the other hand (Fig.~\ref{coexist}b), an isotropic region void of the Chern-Simons charges can coexist with an anisotropic region with charge density $a=a_2$ if the line drawn from $(0,f(T,0))$ to $(a_2,f(T,a_2))$ is tangent to $f(T,a)$ at $a=a_2$. This guarantees mechanical equilibrium.

\subsubsection{$T$-$a$ phase diagram}

\begin{figure}[htb]
  \centering
  \includegraphics[width=10cm]{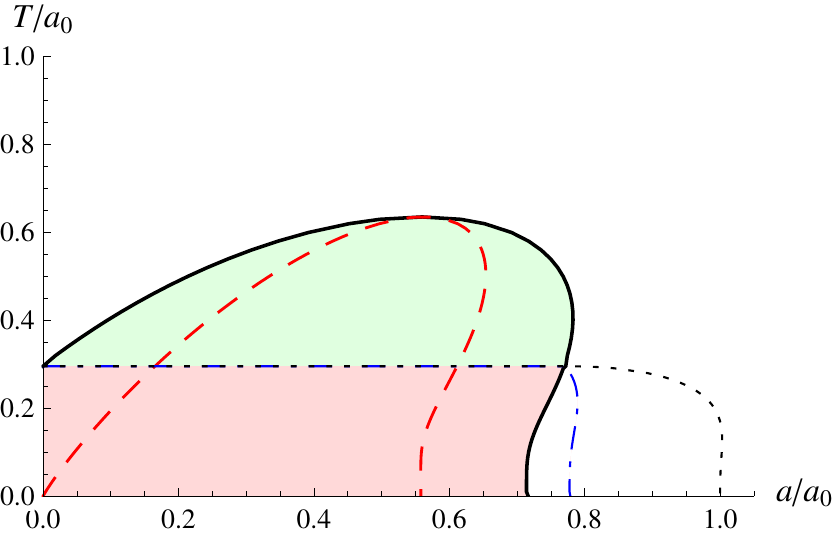}
  \caption{\label{phasediag}The phase diagram. The solid line separates the homogeneous and inhomogeneous (shaded region) phases. In the green shaded part of the phase diagram, the inhomogeneous phase consists of separate anisotropic regions with different values of $a$, whereas in the red shaded part, the plasma contains anisotropic and isotropic regions. The red dashed line indicates the region where the (would-be) homogeneous phase is unstable. The blue dash-dotted line indicates vanishing chemical potential. Inside the line, the pressure anisotropy is oblate, outside it is prolate. 
The dotted line indicates the region where $f(a,T)<f(0,T)$.}
\end{figure}

The phase diagram of the stable phases can now be computed, and is given in Fig.\ \ref{phasediag}. The shaded area indicates the region in phase space where thermodynamic equilibrium is reached in the inhomogeneous phase with a mix of values $a_1$ and $a_2$ given by the boundary values of $a$ of the shaded region at the given temperature. Within the inhomogeneous region, we have indicated with a dashed line the region where the homogeneous phase would be thermodynamically unstable. Outside the dashed line the homogeneous phase is at least metastable.
In the red shaded area, the inhomogeneous phase consists of isotropic and anisotropic (finite Chern-Simons charge density) regions. In the green shaded region the system is, in addition to being inhomogeneous, also anisotropic everywhere.


The type of (pressure) anisotropy of the system (prolate vs.\ oblate) is determined by the sign of the chemical potential $\Phi=\partial f/\partial a$. For $\Phi > 0$, the plasma is prolate, i.e, $p_\parallel>p_\perp$, see Eq.~(\ref{pparallel}). In Fig.~\ref{phasediag}, the boundary in phase space separating prolate plasma from oblate plasma is given by the blue dot-dashed line. Outside that line, the plasma is prolate. Note that in the red shaded part of the phase diagram, the chemical potential does not have a unique value within the plasma: the plasma is composed of isotropic ($\Phi=0$) and oblate ($\Phi<0$) regions. In the green shaded part of the phase diagram, the plasma contains only prolate ($\Phi>0$) regions, which are in chemical equilibrium (same $\Phi$ but different $a$). 

In the homogeneous phase (unshaded region in Fig.\ \ref{phasediag}), the plasma is prolate for the major part of the phase diagram. There is only a narrow region at low temperatures where the plasma is homogeneous and oblate (the unshaded region between the full line and the blue dash-dotted line).

Note that we have chosen to characterise the plasma as oblate, prolate or isotropic only with regard to the ratio of longitudinal and transverse pressures. Intrinsically the plasma is anisotropic whenever $a\not=0$, but it can have isotropic pressure also for nonvanishing $a$, when $\Phi=0$ (along the blue dash-dotted line in Fig.\ \ref{phasediag}).

\subsubsection{Metastable homogeneous phases}
\label{sec:metastableweak}

\begin{figure}[htb]
  \centering
  \includegraphics[width=10cm]{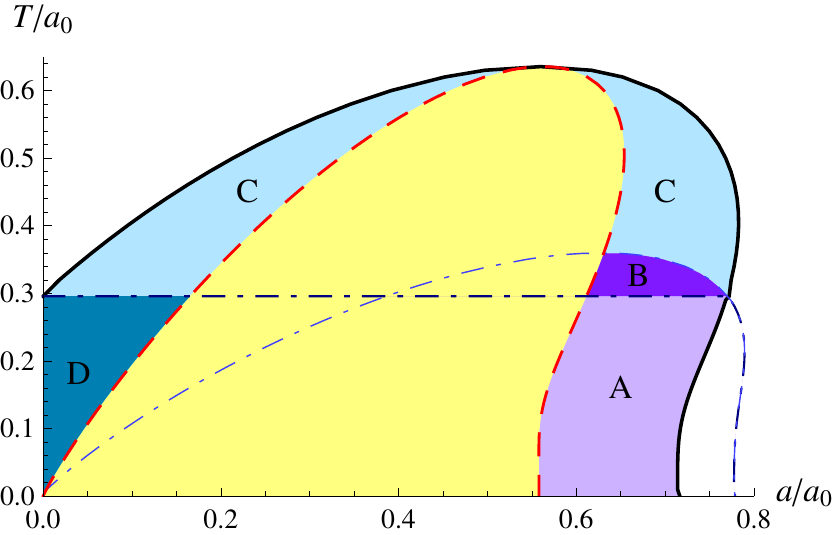}
  \caption{\label{metastablediag}The phase diagram showing different metastable phases labelled A--D with homogeneous Chern-Simons charge distribution (the thermodynamically unstable region is shown in yellow). The light-blue
dash-dotted curve corresponds to $\Phi=0$ in the homogeneous case, with $\Phi>0$ (prolate pressure anisotropy) and $\Phi<0$ (oblate) above and below this line, respectively. Above the straight dark-blue dash-dotted line the inhomogeneous ground state is composed of differently prolate plasma, below this line
the ground state is a mix of isotropic and oblate plasma.}
\end{figure}

In the region of the phase diagram where the inhomogeneous phase is
energetically preferred (in the area below the solid line in Fig.\
\ref{phasediag}), one can distinguish a number of qualitatively different
metastable situations for a homogeneous charge distribution
as shown in Fig.\ \ref{metastablediag}.
In this diagram, the metastable homogeneous phases are labelled A--D while the region where the homogeneous phase is thermodynamically
unstable is colored yellow. For a homogeneous charge distribution
the dividing line between oblate and prolate pressure anisotropy is
given by light-blue dash-dotted curve, whereas for the inhomogenous
phase the dividing line is the straight dark-blue dash-dotted line.
This difference is responsible for the appearance of four different
``decay modes'' of homogeneous metastable phases with density $a$:
\begin{itemize}
\item[(A)]the metastable homogeneous phase is oblate and decays
into a mix of regions that are isotropic ($a_1=0$) and oblate ($a_2>a$)
\item[(B)]the metastable homogeneous phase is oblate and decays
into a mix of prolate regions with different nonvanishing densities
$a_1<a$ and $a_2>a$
\item[(C)]the metastable homogeneous phase is prolate and decays
into a mix of prolate regions with different nonvanishing densities
$a_1<a$ and $a_2>a$
\item[(D)]the metastable homogeneous phase is prolate and decays
into a mix of regions that are isotropic ($a_1=0$) and oblate ($a_2>a$)
\end{itemize}

Note that the metastable homogeneous phases of type A and D,
which like all homogeneous phases are trivially in chemical equilibrium,
decay into inhomogeneous systems that are no longer in chemical equilibrium.

\section{Comparison with holographic infinite-coupling results}

In Ref.~\cite{Mateos:2011ix,Mateos:2011tv} the gravity dual to the
strong-coupling limit of maximally supersymmetric Yang-Mills theory
at infinite color number $N_c$ with Chern-Simons charge
density $a$ has been constructed and worked out in great detail.

In the 5-dimensional bulk, this involves a linear axion field $\chi=az$ and an anisotropic
metric of the form
\begin{align}
 ds^2=\frac{1}{u^2}\Big(-\mathcal{F}(u)\mathcal{B}(u)dt^2+dx^2+dy^2+\mathcal{H}(u)dz^2+\frac{du^2}{\mathcal{F}(u)}\Big),
 \label{eq:metricMT}
\end{align}
which at finite temperature has a regular horizon at some value $u=u_h$
and approaches anti-de Sitter form at the boundary $u=0$.

At nonvanishing $a$, the gauge theory has a trace anomaly, which
divided by $N_c^2$ reads~\cite{Mateos:2011tv}
\be
(\epsilon-2 p_\perp-p_\parallel)/N_c^2 = -\frac{ a^4}{48 \pi^2}\,.
\ee
This is in fact curiously close to the zero-coupling 
(and nonsupersymmetric) result (\ref{tranom0})
which equals $-a^4/(51.2\, \pi^2)$.

When expressed in terms of $a_0$, defined such that $f(0,a_0)=0$ at nonvanishing $a$'s, the result for $f$ at strong coupling is just proportional to the zero-coupling case, which implies that 
$a_\Phi, a_{p_\parallel}, a_{\Phi'}$ are exactly the same as in
Eqs.~(\ref{aPhi0})--(\ref{aPhip0}). Therefore, the phase diagrams
of the two theories when
drawn in units of $a_0$
exactly coincide in the limit of vanishing temperature.

\begin{figure}[tb]
  \centering
  \includegraphics[width=10cm]{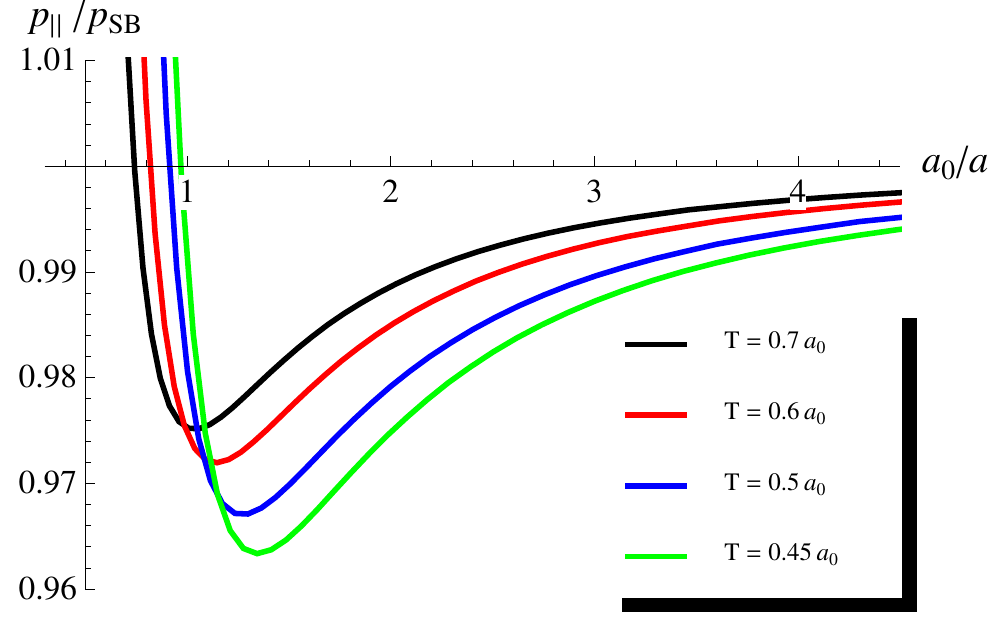}
  \caption{\label{pLplotstrong}Longitudinal pressure as a function of $1/a$ for various temperatures at strong coupling, normalized to Stefan-Boltzmann pressure.}
\end{figure}

At nonzero temperature, there are however significant differences.
At low temperatures ($T\ll a$), the free energy behaves as \cite{Mateos:2011tv}
\be
[f(T,a)-f(0,a)]/N_c^2\sim -0.9\, a^{1/3}T^{11/3}
\ee
while at zero coupling we had obtained $-0.076 a^{1/2} T^{7/2}$
in Eq.\ (\ref{flowT0}).

At high temperatures, the strong coupling result reads \cite{Mateos:2011tv}
\be
[f(T,a)-f(0,a)]/N_c^2=
-\frac{\pi^2 T^4}8 - \frac {a^2T^2}{32} + O(a^4),
\ee
which is to be contrasted with Eq.\ (\ref{fhiT0}), where we had
$-\pi^2 T^4/45+a^2T^2/48-a^3 T/64+O(a^4)$.

The term cubic in a mass parameter and linear in $T$ in the zero-coupling result is typical
of a weakly coupled plasma involving massive quasiparticles; its absence in the
strong-coupling result illustrates that a quasiparticle description
is no longer possible there.

The different sign in the $a^2T^2$ correction implies a different sign for $\Phi=\partial f/\partial a$, which is solely responsible for the pressure anisotropy. 
This means that in the high-temperature limit the
strong-coupling system has oblate pressure anisotropy, whereas
it was prolate at zero coupling.
This also turns out to completely change the structure of the
phase diagram at high temperature.

In Fig.\ \ref{pLplotstrong} we have evaluated numerically the longitudinal
pressure of the strong-coupling theory as a function of $1/a$ for the same set of temperatures (in terms of the scale $a_0$) as in the zero-coupling
theory (Fig.~\ref{pLplot}). In contrast to the latter we now find that
increasing the temperature does not get rid of unstable regions 
in $a$; the homogeneous phase is now always unstable for sufficiently
small $a$ while stable for larger $a$.

\begin{figure}[tb]
  \centering
  \includegraphics[width=7.5cm]{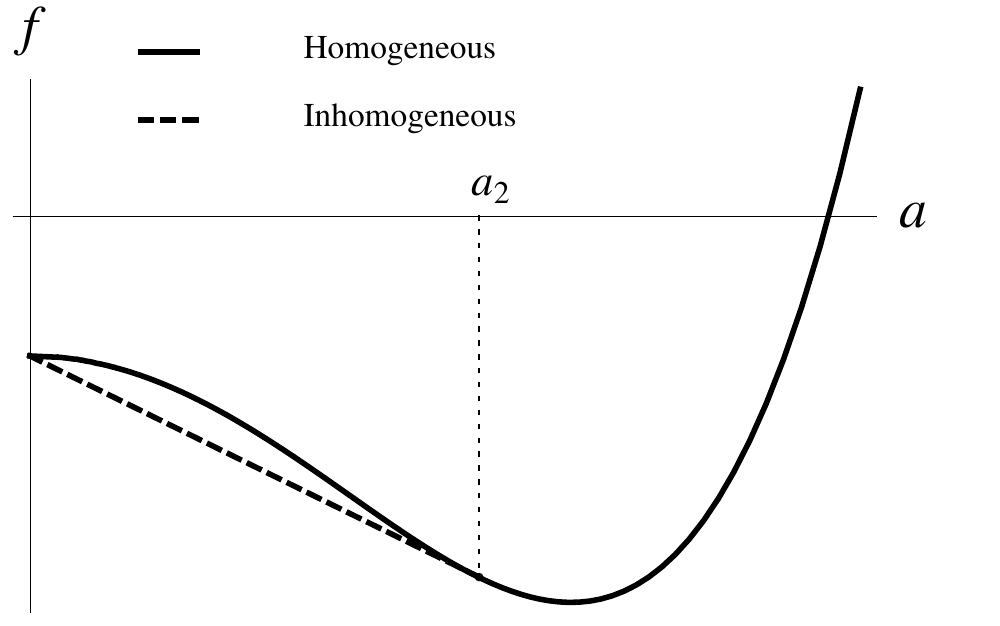}
  \caption{\label{coexiststrong}Condition for coexistence of isotropic and anisotropic regions at strong coupling. The dashed line, giving the free energy of the inhomogeneous phase, is the tangent to $f(T,a)$ at $a=a_2$ and coincides with $f(T,a)$ at $a=0$.}
\end{figure}

\begin{figure}[htb]
  \centering
  \includegraphics[width=10cm]{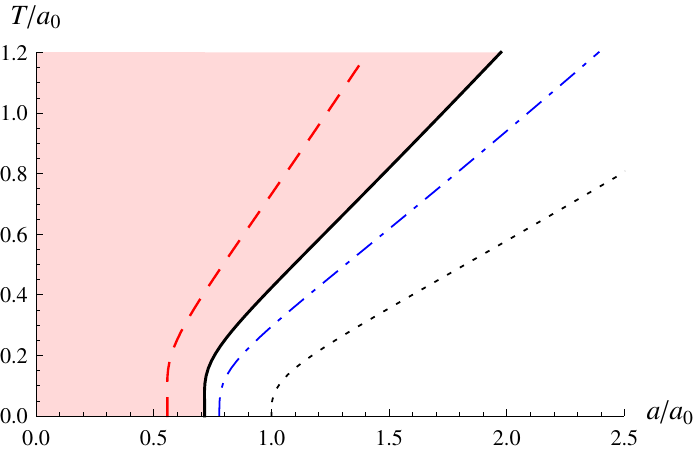}
  \caption{\label{phasediagstrong}The phase diagram at strong coupling. Labeling of the different curves as in Fig.~3.}
\end{figure}

Fig.\ \ref{coexiststrong} shows a typical dependence of $f(a,T)$ on $a$.
Because the curvature of $f$ is negative at $a=0$, but $f$ behaves
as $a^4\ln a$ for sufficiently large $a$, there is always a region in $a$
around $a=0$ where the free energy can be lowered by an inhomogeneous
mix of isotropic ($a=0$) and more strongly anisotropic ($a=a_2$) domains.

In Fig.\ \ref{phasediagstrong} our numerical result for the full phase diagram at strong coupling is shown,
which agrees with the qualitative sketch given in Ref.\ \cite{Mateos:2011tv}%
\footnote{The particular bending of the curves at low temperatures has been
greatly exaggerated in Ref.\ \cite{Mateos:2011tv}, but is qualitatively
correct apart from the fact that all lines become strictly
vertical in the limit $T\to0$.}.
Comparing with the zero-coupling result shown in
Fig.\ \ref{phasediag} we see that at small temperatures
the two phase diagrams are very similar, but
at high temperatures there are significant differences.

In the strong-coupling case, instabilities of
homogeneous phases against the formation of inhomogeneous
structures along the $z$-direction
only occur in the region left to the blue dash-dotted line
where the pressure anisotropy is oblate, but the oblate region
now extends to arbitrarily high temperature.
Between the blue dash-dotted line and the full line
the oblate homogeneous plasma is stable. In the red-shaded region
the energetically preferred inhomogeneous phase always corresponds to a mixture
of isotropic ($a=0$) and more strongly anisotropic regions.
To the left of the red dahed line, the homogeneous phase is
thermodynamically unstable, to the right it is at least metastable.
Only type A of the metastable phases discussed in Sect.\ 
\ref{sec:metastableweak} is realized at strong coupling.

As we have seen above, in the weak-coupling case, sufficiently high temperature
implies prolate pressure aniso\-tropy and stability of
a homogeneous distribution of Chern-Simons charge.
However, at moderate temperature, there are
also domains with prolate anisotropy that are
unstable as well as inhomogeneous phases
with a mixture of prolate plasma with different
nonzero Chern-Simons charge densities (the green-shaded
area in Fig.\ \ref{phasediag}), which
do not occur in the strong-coupling case.

\section{Conclusion}

We have worked out the complete thermodynamics and phase structure
of an Abelian gauge theory at finite temperature that is rendered anisotropic
by a finite density of charges which couple to a 
2+1-dimensional Chern-Simons operator. The results obtained
can be expected to remain qualitatively unchanged also at
small coupling of its non-Abelian version.
In maximally
supersymmetric form the latter can be studied by gauge-gravity duality
in the limit of infinite coupling and color number
\cite{Mateos:2011ix,Mateos:2011tv}.

Because of the resulting nontrivial phase structure, already the Abelian
model studied here is certainly interesting in its own right.
The physics motivation for considering an external
density of charges with anisotropic coupling to gauge bosons
is that it may provide a tractable toy model of
anisotropies occurring in the (nonequilibrium) dynamics of
quark-gluon plasma produced in heavy-ion collision,
which is expected to be radically different
at weak and strong coupling.

In this paper, we have compared the resulting phase diagrams 
of this toy model in the weak and the
strong coupling cases and found stark differences
at high temperatures, but complete agreement in the
limit of zero temperature (when expressed in terms of
the scale $a_0$). The differences are mainly due to
the different signs of the free energy contributions
of  the Chern-Simons charge density $a$ when $a\ll T$.
At weak coupling, $a>0$ introduces anisotropic mass terms
for the gauge bosons which reduce the absolute value
of the free energy compared to the isotropic Stefan-Boltzmann result.
At strong coupling, nonzero $a$ has the opposite effect.
For the same reason, the pressure anisotropy at weak coupling
is prolate in the high-temperature (low $a$) limit, 
while at strong coupling it is oblate.

For $T \lessapprox 0.3 a_0$,
the weak and the strong coupling limits lead to the same
kind of instabilities of homogeneous phases in regions
of their $T$-$a$ phase diagrams corresponding
to oblate pressure anisotropy. There are metastable and
completely unstable regions where it is energetically
favorable to have the charge density $a$ redistributed
along the axis of anisotropy so that a mixture of denser
and empty (isotropic) regions arises in the form of stacks which
are homogeneous in the transverse directions.
The strong coupling results differ from the weak coupling ones
in that the oblate phase
extends to infinite temperature
and so do the inhomogeneous phases appearing within that.
At weak coupling, on the other hand,
the appearance of inhomogeneous phases is restricted to $T$ and $a$ 
both smaller than $a_0$.
Moreover, for moderate temperatures there are also
mixed phases consisting of different prolate plasma; those
phases have no counterpart in the strong coupling limit.

Comparing to the situation of a weakly coupled plasma
with momentum-space aniso\-tropy we note that the instabilities
of the present model are vaguely reminiscent of plasma
instabilities. However, as we have seen, in the anisotropic Chern-Simons
deformed theory they are not associated with
unstable modes in the gauge fields. While this
may point to limitations of this model as a model for
the nonequilibrium situation of an anisotropic quark-gluon plasma,
it also provides opportunities for further studies.
As mentioned in the introduction,
in the strong-coupling limit also several transport coefficients
of relevance to heavy-ion phenomenology have been worked out
already in the present model of an anisotropic plasma. 
It might be interesting to also obtain the corresponding
results at weak coupling.
Because of the absence of tachyonic modes
in the gauge boson spectrum, such calculations are in principle feasible,
while the hard-loop effective theory of a weakly coupled plasma
with momentum-space anisotropy typically suffers from
nonintegrable singularities \cite{Romatschke:2006bb,Baier:2008js,Carrington:2009vm}.

\subsection*{Acknowledgments}

This work was supported by the Austrian Science
Fund FWF, project no. P22114.

\appendix

\section{Computations}
\label{app:sec-1}

In the presence of the source $j(z)=a$, the path integral in Eq.~(\ref{pathint}) can be carried out immediately with the result
\begin{eqnarray}
  \mathcal{Z}(T,a) & = & e^{-\beta V \Omega}\frac{\det_K(\Delta^{-1}(K))}{\sqrt{\det_{\mu\nu,K} (D^{-1}_{\mu\nu}(K))}},
\end{eqnarray}
where the inverse photon and ghost propagators are given in momentum space by
\begin{eqnarray}
  D^{-1}_{\alpha\beta}(K) & = & \beta^2 K^2 \delta_{\alpha\beta}, \\
  D^{-1}_{ij}(K) & = & \beta^2\left[K^2\delta_{ij} - a\epsilon_{ijk}k_k\right], \\
  D^{-1}_{i\alpha}(K) & = & D^{-1}_{\alpha i}(K) \, = \, 0, \\
  \Delta^{-1}(K) & = & \beta^2 K^2.
\end{eqnarray}
The determinant of the inverse photon propagator over the Lorentz indices is given by
\begin{eqnarray}
  \|D_{\mu\nu}^{-1}\| & = & (\beta^2K^2)^{1-2\epsilon} \beta^6 K^2\left[K^4 + (K^2-k_\parallel^2) a^2\right] \nonumber \\
  & = & (\beta^2)^{4-2\epsilon} (K^2)^{2-2\epsilon}\prod_{\pm}\left(K^2 + \frac{a^2 \pm \sqrt{a^4+4a^2 k_\parallel^2}}{2}\right).
\end{eqnarray}
It can immediately be seen that the ghost contribution cancels the contribution from the two $a$-independent photon modes. The logarithm of the partition function can thus be written as
\begin{eqnarray}
  \ln \mathcal{Z}(T,a) & = & -\frac{\beta V}{2}\sumint{K}\sum_{\pm}\ln \left(\beta^2 k_0^2 + \beta^2 \big(k_\perp^2 + M_\pm^2(k_\parallel)\big)\right) - \beta V \Omega, 
\end{eqnarray}
where
\begin{eqnarray}
  M_\pm^2(k_\parallel) & = & k_\parallel^2 + \frac{a^2 \pm \sqrt{a^4+4a^2 k_\parallel^2}}{2},
\end{eqnarray}
or, equivalently,
\be
M_\pm(k_\parallel) 
=  \sqrt{k_\parallel^2 + a^2/4} \pm \frac{a}{2}. 
\ee

Note that in Minkowski space ($k_0\to i\omega$) we have photon modes that for nonzero $k_\parallel$ split into modes with $\omega_+^2>\mathbf{k}^2=k_\perp^2+k_\parallel^2$ and ones with $\omega_-^2<\mathbf{k}^2$. However, also the latter is non-tachyonic since $\omega_-^2\ge0$.

Using
\begin{eqnarray}
  \sum_{n=-\infty}^\infty\partial_{x^2}\ln(4\pi^2 n^2 + x^2) & = & \sum_{n=-\infty}^\infty\frac{1}{4\pi^2 n^2 + x^2} \, = \, \frac{1}{2x}\left[1 + \frac{2}{e^x-1}\right] \nonumber \\
  & = & \partial_{x^2}\left[x + 2 \ln\left(1-e^{-x}\right)\right] \\
  \Rightarrow \sum_{n=-\infty}^\infty\ln(4\pi^2 n^2 + x^2) & = & x + 2 \ln\left(1-e^{-x}\right) + \mathrm{constant}, 
\end{eqnarray}
we get for the Helmholtz free energy density $f = -T/V \ln\mathcal{Z}$ (with the constant absorbed into the cosmological constant)
\begin{eqnarray}
  f(T,a) & = & \Omega + T\sum_{\pm}\int_k\left[\frac{1}{2}\beta\omega_\pm + \ln\left(1-e^{-\beta\omega_\pm}\right)\right],
\end{eqnarray}
where
\begin{eqnarray}
  \omega_\pm^2 & = & k_\perp^2 + M_\pm^2(k_\parallel).
\end{eqnarray}

\subsection{$T=0$ contribution}
\label{app:sec-1.1}

The $T=0$ contribution to the free energy density is given by
\begin{eqnarray}
  f(0,a) & = & \Omega + \frac{1}{2}\int_k\left[\omega_+(k_\perp,k_\parallel)+\omega_-(k_\perp,k_\parallel)\right] \\
  & = & \Omega_B(\Lambda_\perp,\epsilon) + \frac{1}{4\pi}\int_{k_\parallel} \int_0^{\Lambda_\perp} dk_\perp k_\perp\left[\sqrt{k_\perp^2+M_+^2(k_\parallel)}+\sqrt{k_\perp^2+M_-^2(k_\parallel)}\right], \nonumber \end{eqnarray}
where we have introduced a UV regulator for the integration over the transverse momenta and anticipated that the zero-loop contribution (which is just $\Omega$) will need to be renormalized. The integral over the transverse momentum can be carried out easily, giving
\begin{eqnarray}
  \int_0^{\Lambda_\perp} dk_\perp k_\perp\sqrt{k_\perp^2+M_\pm^2(k_\parallel)} & = & \int_0^\infty dk k\left[\sqrt{k^2+M_\pm^2(k_\parallel)} - k - \frac{M_\pm^2(k_\parallel)}{2k}\right] \nonumber \\
  & & + \int_0^{\Lambda_\perp}dk \left(k^2 + \frac{1}{2}M_\pm^2(k_\parallel)\right) + \mathcal{O}(1/\Lambda_\perp) \\
  & = & -\frac{1}{3}M_\pm^3(k_\parallel) + \frac{1}{3}\Lambda_\perp^3 + \frac{1}{2}M_\pm^2(k_\parallel)\Lambda_\perp + \mathcal{O}(1/\Lambda_\perp).\nonumber
\end{eqnarray}
Inserting this into $f(0,a)$ above, dropping terms that vanish as the regulators are removed, we get
\begin{eqnarray}
  f(0,a) & = & -\frac{1}{12\pi}\int_{k_\parallel} \left[M_+^3(k_\parallel)+ M_-^3(k_\parallel)\right] + \frac{\Lambda_\perp}{4\pi}\int_{k_\parallel} \left(2k_\parallel^2+a^2\right) + \Omega_B(\Lambda_\perp,\epsilon) \nonumber\\
  & = & -\frac{1}{12\pi}\int_{k_\parallel} \left[M_+^3(k_\parallel)+ M_-^3(k_\parallel)\right] + \Omega_B(\Lambda_\perp,\epsilon).
\end{eqnarray}
Extracting the large-$k_\parallel$ behaviour, the first term can be written as
\begin{eqnarray}
  \sum_{\pm}\int_{k_\parallel} M_\pm^3(k_\parallel)& = & \frac{1}{\pi}\int_{0}^\infty dk_\parallel\left[M_+^3(k_\parallel)+ M_-^3(k_\parallel) - 2 k_\parallel^3 - \frac{9}{4}a^2k_\parallel - \frac{15 a^4}{64\sqrt{k_\parallel^2+a^2}}\right] \nonumber \\
  & & +\int_{k_\parallel} \left[2k_\parallel^2 + \frac{9}{4}a^2k_\parallel + \frac{15a^4}{64\sqrt{k_\parallel^2+a^2}}\right] + \mathcal{O}(\epsilon) \nonumber \\
  & = & \frac{c\, a^4}{\pi} + \frac{15a^4}{128 \pi}\left(\frac{1}{\epsilon} - \ln\frac{a^2}{\bar{\Lambda}^2}\right) + \mathcal{O}(\epsilon),
\end{eqnarray}
where
\begin{eqnarray}
  c & = & \int_0^\infty dx \left[\left(x^2+\frac{1}{2}(1+\sqrt{1+4x^2})\right)^{3/2}+\left(x^2+\frac{1}{2}(1-\sqrt{1+4x^2})\right)^{3/2}\right. \nonumber \\
    & & \quad \quad \left.-2 x^3 - \frac{9}{4}x - \frac{15}{64\sqrt{1+x^2}}\right] = 0.29136\ldots  
\end{eqnarray}
The free energy density at $T=0$ is then
\begin{eqnarray}
  f(0,a) & = & -\frac{c\,a^4}{12\pi^2} - \frac{5 a^4}{512 \pi^2}\left(\frac{1}{\epsilon} - \ln\frac{a^2}{\bar{\Lambda}^2}\right) + \Omega_B(\Lambda_\perp,\epsilon) \nonumber\\
  & \equiv & -\frac{c\,a^4}{12\pi^2} + \frac{5 a^4}{256 \pi^2}\ln\frac{a}{\bar{\Lambda}} + \Omega(\bar{\Lambda}),
\end{eqnarray}
where $\Omega(\bar{\Lambda})$ is the renormalized cosmological constant in the $\ms$-scheme. 
\subsection{Finite $T$ contribution}
\label{app:sec-1.2}

The finite-$T$ contribution is given by
\begin{eqnarray}
  f(T,a)-f(0,a) & = & T\sum_{\pm}\int_k\ln\left(1-e^{-\beta \omega_\pm}\right).
\end{eqnarray}
The integral over the transverse momenta can be carried out immediately using standard integrals. Noting that
\begin{eqnarray}
  2T \frac{\partial}{\partial k_\perp^2}\ln\left(1-e^{-\beta\omega_\pm}\right) & = & \frac{1}{\omega_\pm}\frac{1}{e^{\beta\omega_\pm}-1},
\end{eqnarray}
we find
\begin{eqnarray}
  T \int\frac{d^2k_\perp}{(2\pi)^2} \ln\left(1-e^{-\beta \omega_\pm}\right) & = & -\frac{3 T^3}{2 \pi}h_4\left(M_\pm(k_\parallel)/T,0\right),
\end{eqnarray}
where the function $h_n(y,r)$ is defined by
\begin{eqnarray}
  h_n(y,r) & = & \frac{1}{\Gamma(n)}\int_0^\infty dx \frac{x^{n-1}}{\sqrt{x^2+y^2}}\frac{1}{e^{\sqrt{x^2+y^2}-ry}-1}.
\end{eqnarray}
Their properties are discussed in detail in Ref.~\cite{Haber:1981tr}. 
In particular,
\begin{eqnarray}
  \frac{\partial}{\partial y}h_{n+1}(y,0) & = & -\frac{y}{n}h_{n-1}(y,0), \\
  h_2(y,0) & = & -\ln\left(1-e^{-y}\right), \\
  h_4(y,0) & = & \frac{1}{3}\left[y\, \mathrm{Li}_2\left(e^{-y}\right) + \mathrm{Li}_3\left(e^{-y}\right)\right].
\end{eqnarray}
We now get for the integral over the longitudinal momentum,
\begin{eqnarray}
  \int_{-\infty}^\infty \frac{dk_\parallel}{2\pi}h_4\left(M_\pm/T,0\right) & = & \frac{1}{\pi}\int_0^\infty dk_\parallel\, h_4\left(M_\pm/T,0\right) \nonumber \\
  & = & \frac{1}{3\pi T^2}\int_0^\infty dk_\parallel\, k_\parallel\, M_\pm \frac{dM_\pm}{dk_\parallel} h_2\left(M_\pm/T,0\right) \\
  & = & -\frac{T}{3\pi}\int_0^\infty dx\, x^2\left(1 \pm \frac{y}{\sqrt{x^2+y^2}}\right)\ln\left(1-e^{-\sqrt{x^2+y^2}\, \mp y}\right), \nonumber
\end{eqnarray}
where we have denoted $y=a/(2T)$. The finite-$T$ contribution is thus given by the integral
\begin{equation}
    f(T,a)-f(0,a) \, = \, \frac{T^4}{2\pi^2}\sum_{\pm}\int_0^\infty dx\, x^2\left(1 \pm \frac{y}{\sqrt{x^2+y^2}}\right)\ln\left(1-e^{-\sqrt{x^2+y^2}\, \mp y}\right)\,.
\end{equation}

\subsubsection{High- and low-$T$ expansions}
\label{app:sec-1.2.1}

Even though we shall study the thermodynamic properties of this system numerically, it is instructive to calculate the analytic high- and low-$T$ approximations to the free energy as well. As derived above, the finite-$T$ contribution is given by
\begin{eqnarray}
  f(T,a)-f(0,a) & = & -\frac{4 T^4}{\pi^2}\left[h_5(y,1)+h_5(y,-1)\right] + \frac{T^4}{2\pi^2} I(y,1),
\end{eqnarray}
where
\begin{eqnarray}
  I(y,r) & = & \int_0^\infty dx \frac{yx^2}{\sqrt{x^2+y^2}}\left[\ln\left(1-e^{-\sqrt{x^2+y^2}-ry}\right) - \ln\left(1-e^{-\sqrt{x^2+y^2}+ry}\right) \right].\quad
\end{eqnarray}
We immediately find that
\begin{eqnarray}
  I(y,0) & = & 0, \\
  \frac{\partial I}{\partial r} & = & 2y^2 \left[ h_3(y,r) + h_3(y,-r) \right]
\end{eqnarray}
and thus
\begin{eqnarray}
  I(y,1) & = & 2y^2\, \int_0^1 dr \left[h_3(y,r)+h_3(y,-r)\right].
\end{eqnarray}
The high-$T$ (small-$y$) expansions of $h_n(y,r) + h_n(y,-r)$ have been computed 
in Ref.~\cite{Haber:1981tr}, with the results
\begin{eqnarray}
  h_5(y,1) + h_5(y,-1) & = & \frac{\pi^4}{180} + \frac{\pi^2 y^2}{48} +\frac{y^4}{64}\left(\ln\frac{y}{4\pi} + \gamma_E + \frac{7}{12} \right) + \mathcal{O}(y^6), \\
  h_3(y,r) + h_3(y,-r) & = & \frac{\pi^2}{6} - \frac{\pi y}{2}\sqrt{1-r^2} - \frac{y^2}{4}\left(\ln\frac{y}{4\pi} +\gamma_E - \frac{1}{2} + r^2 \right) + \mathcal{O}(y^4)\qquad
\end{eqnarray}
Putting everything together, and setting $y=a/(2 T)$ we get
\begin{equation}
    f(T,a)-f(0,a) \, = \, -\frac{\pi^2 T^4}{45} + \frac{a^2 T^2}{48} - \frac{a^3 T}{64} - \frac{5 a^4}{256\pi^2}\left(\ln\frac{a}{8\pi T} + \gamma_E - \frac{1}{60} \right) + \mathcal{O}(a^6T^{-2})
\end{equation}
for the high-$T$ expansion. On the other hand, at low temperatures (high-$y$ limit) we have
\begin{eqnarray}
h_5(y,1)+h_5(y,-1) & = & \frac{1}{64}\sqrt{\frac{\pi y}{2}}\Big(8\zeta(5/2) y + 15\zeta(7/2) + \mathcal{O}(1/y)\Big) \\
h_3(y,r)+h_3(y,-r) & = & \frac{1}{2}\sqrt{\frac{\pi y}{2}}\left[\mathrm{Li}_{3/2}(e^{(r-1)y}) +\mathrm{Li}_{3/2}(e^{-(r+1)y}) \right. \nonumber \\ 
& & \left. + \frac{3}{8y}\left(\mathrm{Li}_{5/2}(e^{(r-1)y}) +\mathrm{Li}_{5/2}(e^{-(r+1)y}) \right) + \mathcal{O}(1/y^2) \right] \\
\Rightarrow \, I(y,1) & = & \frac{1}{8}\sqrt{\frac{\pi y}{2}}\Big(8\zeta(5/2) y + 3\zeta(7/2) + \mathcal{O}(1/y) \Big)
\end{eqnarray}
Combining the results gives us
\begin{equation}
    f(T,a)-f(0,a) \, = \, - \frac{3\zeta(7/2)}{8\pi^{3/2}}a^{1/2}T^{7/2} + \mathcal{O}(a^{-1/2})
\end{equation}
in the low-$T$ limit.

\bibliographystyle{JHEP}

\bibliography{athermo}

\end{document}